\documentclass[12pt]{article}
\usepackage[utf8]{inputenc}
\usepackage{amsmath}
\usepackage{amsfonts}
\usepackage{amssymb}
\usepackage{epsfig}
\usepackage{epstopdf}
\usepackage{indentfirst}
\usepackage{geometry}
\usepackage{graphicx}
\usepackage{caption}
\usepackage{subfig}
\usepackage[countmax]{subfloat}
\usepackage{rotating}
\usepackage{hyperref}
\usepackage{lscape}
\usepackage[numbers,sort&compress]{natbib}
\usepackage{tabularx}
\usepackage{setspace}
\usepackage{enumerate}
\usepackage{lineno}
\usepackage[section]{placeins}

\numberwithin{equation}{section}
\def\bA{{\boldsymbol A}}

\def\bP{{\boldsymbol P}}
\def\bQ{{\boldsymbol Q}}

\def\bP{{\boldsymbol P}}

\def\b1{{\boldsymbol 1}}
\def\b0{{\boldsymbol 0}}

\def\bbI{{\mathbb I}}

\def\cP{{\mathcal P}}

\def\cL{{\mathcal L}}

\graphicspath{%
    {converted_graphics/}
    {/}
}

\begin{document}

\title{On spooky action at a distance\\ and conditional probabilities}
\author{Henryk Gzyl\\
Center for Finance, IESA School of Business\\
{henryk.gzyl@iesa.edu.ve}
} 

\date{}
 \maketitle

\baselineskip=1.5\baselineskip \setlength{\textwidth}{6in}

\begin{abstract}
Here we establish that there is no spooky action at a distance. The expression was coined by Einstein to describe the situation in which one can know the exact value of an observable, when the value of another related observable is measured.

Our argument is based on the relationship between the von Neumann post-measurement states and (quantum) conditional probabilities. What happens is that when two random events are dependent with respect to a probability distribution, observing one of them, changes both the available class of microstates of the joint system and the underlying probability distributions. The new probability distribution depends becomes a conditional 
probability distribution.

To bring the relationship forth, we consider a quantum system with two components and a binary observable (a projector) acting on each. To emphasize the predictive logic, we also consider and a classical systems with two dependent binary random variables, and verify that predictions when information is available obey the same logic. 
\end{abstract}

\textbf{Keywords}: Action at a distance, prediction, conditional probabilities as predictors, Measurement and state change.

\begin{spacing}{0.1}
   \tableofcontents
\end{spacing}

\section{Introduction and preliminaries} 
About 90 years ago Einstein, Podolsky and Rosen published a paper, \cite{EPR} in which they proposed a criticism of quantum mechanics as a model for the intrinsic randomness in microscopic phenomena. The core of their argument consists in considering a system with two components, in an initial state in which the variables used to characterize the state (in their case the momenta of the individual particles) were not statistically independent. In current terminology, the state was entangled. 
 They argued that when one of the momenta was measured, since the system was prepared in a state of known total momentum, the momentum of the other particle was known. They claimed that this contradicted the Heisenberg uncertainty principle, for then one could measure the position of the second particle, and this measurement does not have infinite variance as required by the Heisenberg uncertainty principle.

The terms ``entangled'' and ``entanglement; had already appeared in Schr\"{o}dinger's \cite{Sch}, mentioned in a way that suggests statistical non-independence, but was not precisely defined for today's standards. By the way, it is in that same work that the famous cat example came up. Two references at an expository level are \cite{TWD} or \cite{Sieg}. See as well \cite{HHH} for detailed mathematical formulations of the notion of entanglement, especially in the Heisenberg representation. Throughout this expos\'e, when we talk about entangled states, we mean entangled in the Schr\"{o}dinger representation. 

About 10 years later, in further discussing with Max Born the notion of physical reality and the use of quantum mechanics to understand it,  Einstein wrote: \textit{I cannot seriously believe in it because the theory cannot be reconciled with the idea that physics should represent a reality in time and space, free from spooky action at a distance}, \cite{EBL}. That phrase ``spooky action at a distance'' is the title of an expository paper  about entanglement, see \cite{Gu2}, which inspired this work.

The aim of this work is twofold. First, we lay a bridge between quantum measurement theory and quantum classical probabilistic prediction theory. Then we use this to explain why there is no spooky action at a distance.

In the Schr\"{o}dinger representation, the notion of entanglement is the analogue to the classical notion of statistical dependence. This conceptual equivalence may be of interest in current applications of quantum mechanics, as in quantum computing, say. See \cite{NC} for example. In this work, the notions of measurement and von Neumann post measurement state, \cite{J,LeB}, \ play an essential role. For an application to understand why the EPR critique of quantum mechanics is not valid, consider \cite{Gz}.

The paper is organized as follows. In the remainder of this section we recall some basic notations along the lines of \cite{Gu2}. In section 2, we reconsider his example and prove that prediction in the von Neumann post-measurement state amounts to prediction with respect to quantum conditional probabilities as introduced in \cite{Gu1}. Actually, in our setup, the technicalities described in \cite{Gu1} do not play a role due to the commutativity of the operators involved. Such commutativity renders the classical and quantum conditional probabilities identical.

Then, in Section 3, we particularize the setting of section 2 to explain why there is no spooky action at a distance. For that, starting from any initial state of the composite system, we prepare a state in which when subsystem 1 the event $\bP$ occurs, then in the subsystem 2 the event only $\bQ^c$ can occur, and when in subsystem 1 the event $\bP^c$ occurs, then in subsystem 2 we must observe the event $\bQ.$ This prior information that must be taken into account when making further predictions. In this set up, suppose that we observe that in subsystem 2, the event $\bQ$ happened. To make further predictions about the system, we form the von Neumann post measurement state, and verify that $\bP$ cannot happen but $\bP^c$ happens with probability 1. That is, it becomes a certain event. 
Nothing is transmitted from one component to another, it is all in the predictive logic. 

After that, in Section 4 we add a few comments about classical statistical independence and its counterpart in the quantum formalism. In the last section 

The basic building block consists of a (complex) Hilbert space $H,$ in which the scalar product between to vectors $\psi$ and $\phi$ is denoted by $\langle\psi,\phi\rangle,$ and $|\langle\psi,\phi\rangle|^2$ is interpreted as the probability of the system being in state $\psi$ when it is  in state $\phi.$ The probabilistic interpretation requires states to be represented by normalized vectors, that is $\|\psi\|^2=|\langle\psi,\psi\rangle|^2=1.$ Observables, (the analogue of classical random variables), are represented by self-adjoint (with respect to the given scalar product), bounded operators.
Their collection is denoted by $\cL(H).$

In the quantum mechanical formalism, events are represented by projection operators $\bP,$ that is, self-adjoint operators such that $\bP^2=\bP,$ and the probability of observing such event when the system is in stated $\psi$ is given by 
\begin{equation}\label{qp1}
\cP_\psi(\bP)= \langle\psi,\bP\psi\rangle.
\end{equation}
And, after the observation of an event $\bP$ is made, then the post-observation state of the system is
\begin{equation}\label{pos1}
\psi' = \frac{\bP\psi}{\|\bP\Psi\|}
\end{equation}
Where, $\|\bP\Psi\|=\langle\psi,\bP\psi\rangle^{1/2}.$ To finish the list of preliminaries, when describing a composite system, whose quantum states are vectors in the Hilbert spaces $H_1$ and $H_2,$ one forms the tensor product space $H=H_1\otimes H_2,$ in which the scalar product of two vectors $\psi_1\otimes\psi_2$ and $\phi_1\otimes\phi_2,$ in Dirac's bra-ket notation, is given by 
$$\langle\psi_1\otimes\psi_2,\phi_1\otimes\phi_2\rangle = \langle\psi_1,\phi_1\rangle\langle\psi_2,\phi_2\rangle.$$
If $\bA_1$ is an observable of the first subsystem, to be proper when thinking of it as an observable of the joint system we should write it as $\bA_1\otimes\bbI_2,$ where $\bbI_2$ is the identity operator in $H_2.$ Such tensor product operator acts on the product state $\psi\otimes\phi$ as follows: 
$$(\bA_1\otimes\bbI_2)\psi\otimes\phi=\bA_1\psi\otimes\phi.$$
A similar comment applies for operators acting on the second subsystem. After this preamble, we move onto the next topic. 

\section{Prediction after a measurement and conditional probabilities}
Here we consider a variation on the example in Section 2 of \cite{Gu2}. 
As initial, entangled state we consider $\eta=a\psi\otimes\phi+b\phi\otimes\psi,$ where
$0<|a|^2, |b|^2<1$ and $|a|^2+|b|^2=1,$ and we suppose that $\langle\psi,\phi\rangle=0.$ Consider events represented by projectors $\bP,$ $\bP^c=\bbI_1-\bP$ on $H_1$ and $\bQ$ and $\bQ^c=\bbI_2-\bQ$ on $H_2.$  Let us now compute the joint probability distribution of these events. 

\begin{gather}
\cP_\eta(\bP,\bQ) = \langle\eta,\bP\bQ\eta\rangle = \;\;\nonumber\\
|a|^2\langle\psi,\bP\psi\rangle\langle\phi,\bQ\phi\rangle+\bar{a}b\langle\phi,\bP\psi\rangle\langle\phi,\bQ\psi\rangle+a\bar{b}\langle\phi,\bP\psi\rangle\langle\phi,\bQ\psi\rangle+|b|^2\langle\phi,\bP\phi\rangle\langle\psi,\bQ\psi\rangle.\;\;\;\label{1,1}\\
\cP_\eta(\bP,\bQ^c) = \langle\eta,\bP\bQ^c\eta\rangle = \;\;\nonumber\\
|a|^2\langle\psi,\bP\psi\rangle\langle\phi,\bQ^c\phi\rangle+\bar{a}b\langle\phi,\bP\psi\rangle\langle\phi,\bQ^c\psi\rangle+a\bar{b}\langle\phi,\bP\psi\rangle\langle\phi,\bQ^c\psi\rangle+|b|^2\langle\phi,\bP\phi\rangle\langle\psi,\bQ^c\psi\rangle.\;\;\;\label{1,2}\\
\cP_\eta(\bP^c,\bQ) = \langle\eta,\bP\bQ\eta\rangle = \;\;\nonumber\\
|a|^2\langle\psi,\bP^c\psi\rangle\langle\phi,\bQ\phi\rangle+\bar{a}b\langle\phi,\bP^c\psi\rangle\langle\phi,\bQ\psi\rangle+a\bar{b}\langle\phi,\bP^c\psi\rangle\langle\phi,\bQ\psi\rangle+|b|^2\langle\phi,\bP^c\phi\rangle\langle\psi,\bQ\psi\rangle.\;\;\;\label{2,1}\\
\cP_\eta(\bP^c,\bQ^c) = \langle\eta,\bP^c\bQ^c\eta\rangle = \;\;\nonumber\\
|a|^2\langle\psi,\bP^c\psi\rangle\langle\phi,\bQ^c\phi\rangle+\bar{a}b\langle\phi,\bP^c\psi\rangle\langle\phi,\bQ^c\psi\rangle+a\bar{b}\langle\phi,\bP^c\psi\rangle\langle\phi,\bQ^c\psi\rangle+|b|^2\langle\phi,\bP^c\phi\rangle\langle\psi,\bQ^c\psi\rangle.\;\;\;\label{2,2}
\end{gather}

Now, if we use the assumption that $\langle\psi,\phi\rangle=0,$ a similar but simpler computation yields:
\begin{gather}
\cP_\eta(\bP)=\langle\eta,\bP\eta\rangle= |a|^2\langle\psi,\bP\psi\rangle+|b|^2\langle\phi,\bP\phi\rangle\;\;\label{mp1}\\
\cP_\eta(\bQ)=\langle\eta,\bQ\eta\rangle= |a|^2\langle\phi,\bQ\phi\rangle+|b|^2\langle\psi,\bQ\psi\rangle\;\label{mp2}
\end{gather}
Notice that the following consistency relationship hold:
\begin{gather}
\cP_\eta(\bP) = \cP_\eta(\bP\bQ) + \cP_\eta(\bP\bQ^c)\;\;\;\label{mar1}\\
\cP_\eta(\bQ) = \cP_\eta(\bP\bQ) + \cP_\eta(\bP^c\bQ)\;\;\;\label{mar2}
\end{gather}
Skipping a number of technicalities (explained in \cite{Gu1}), since in our case the operators $\bP$ and $\bQ$ commute because they operate on different components of the product space,  the notion of classical and quantum conditional probabilities coincide and we have, for example, that the quantum conditional probability of $\bP$ (or of $\bP^c$) given that $\bQ$ is observed to occur, is:

\begin{gather}
\cP_\eta(\bP|\bQ) = \frac{\cP_\eta(\bP\bQ)}{\cP_\eta(\bQ)} \;\;\label{condp1}\\
\cP_\eta(\bP^c|\bQ) = \frac{\cP_\eta(\bP^c\bQ)}{\cP_\eta(\bQ)} \;\;\label{condp2}
\end{gather}
On account of \eqref{mar1}, clearly $\cP_\eta(\bP|\bQ)+\cP_\eta(\bP^c|\bQ)=1.$ That is, the conditional probability distribution is a true probability distribution. Also, we can rewrite \eqref{mar1} and \eqref{mar2} as
\begin{gather}
\cP_\eta(\bP) = \cP_\eta(\bP|\bQ)\cP_\eta(\bQ) + \cP_\eta(\bP|\bQ^c)\cP_\eta(\bQ^c)\;\;\;\label{mar1.1}\\
\cP_\eta(\bQ) = \cP_\eta(\bQ|\bP)\cP_\eta(\bP) + \cP_\eta(\bQ|\bP^c)\cP_\eta(\bP^c)\;\;\;\label{mar2.2}
\end{gather}
And similarly for $\bP^c$ and $\bQ^c,$ further stressing the analogy to the classical setup.

Next, we show that \eqref{condp1} or \eqref{condp2} are the probabilities of the events $\bP$ or $\bP^c$ in the state $\eta_Q$ obtained from $\eta$ according to \eqref{pos1} if the event $\bQ$ is observed. To begin with, note that $\|\bQ\eta\|$ is given by \eqref{mar2}, and that
\begin{equation}\label{pos2}
\eta_Q = \frac{\bbI_1\otimes\bQ\eta}{\|\bQ\eta\|}=\frac{1}{\|\bQ\eta\|}\big(a\psi\otimes\bQ\phi+b\phi\otimes\bQ\psi).
\end{equation}
We leave it for the reader to do the arithmetic and verify that
\begin{gather}
\langle\eta_Q,\bP\eta_Q\rangle = \cP_\eta(\bP|\bQ)\;\;\; \label{corresp1}\\
\langle\eta_Q,\bP^c\eta_Q\rangle = \cP_\eta(\bP^c|\bQ)\;\;\; \label{corresp2}
\end{gather}
That is, the probability of the event $\bP$ in the state $\eta_{\bQ}$ is the (quantum) conditional probability of $\bP$ given $\bQ.$ This is the core of the correspondence between prediction with the von Neumann post measurement state and the prediction with the quantum conditional probabilities.

We close this section with a methodological comment. If we are interested only in the observation of the events $\bP$ and $\bQ$ in each subsystems, since
$$ \bP\otimes\bQ+\bP^c\otimes\bQ+\bP\otimes\bQ^c+\bP^c\otimes\bQ^c=\bbI$$ 
we could have considered a state like:
$$\eta =\alpha(P,Q)\bP\otimes\bQ\Psi+\alpha(P^c,Q)\bP\otimes\bQ\Psi+\alpha(P,Q^c)\bP\otimes\bQ^c\Psi+\alpha(P^c,Q^c)\bP^c\otimes\bQ^c\Psi$$
for some arbitrary choice of $\Psi$ and any coefficients such that $\|\eta\|=1.$
This is clear since in the basis $\bP\otimes\bQ\Psi, \bP^c\otimes\bQ\Psi, \bP\otimes\bQ^c\Psi, \bP^c\otimes\bQ^c\Psi$ are orthogonal for any $\Psi.$  This would have covered all possible choices of initial state, and the analysis and the conclusion obtained above would carry through. 

\section{No spooky action at a distance}
Next, using the notations in the previous section, we imitate the set up in \cite{EPR} to explain why there is no spooky action at a distance. We start from an arbitrary state $\eta$ of the joint system, and by appropriate measurement, we project the state onto a state $\eta$ in which only $\bP$ and$\bQ^c$ or $\bP^c$ and $\bQ,$  occur simultaneously. Represent it as:
\begin{equation}\label{spc1}
\eta_s = \frac{\Pi_s\eta}{\sqrt{\langle\eta|\Pi_s|\eta\rangle}}\;\;\;\mbox{with}\;\;\;\Pi_s=P\otimes Q^c+P^c\otimes Q.
\end{equation}
In the current setup, the spooky action that bothered Einstein refers to the fact that in such a state, if for example, one observes that $\bQ$ happened in the second subsystem, then the probability of $\bP$ being observed in the other subsystem must be zero. Or if we observe that $\bQ^c$ happens in the second subsystem, then we know that $\bP$ happens with probability $1.$ That is we do not have to observe because we know that the event $\bP$ is certain. Let us verify that the predictive logic confirms that. As in \eqref{pos2} denote by $\eta_Q$ the post-measurement state obtained after observing that the second subsystem is in state $\bQ.$ By definition: 

$$\eta_Q = \frac{\bbI_1\otimes\bQ\eta_s}{\|\bQ\eta_s\|}=\frac{1}{\|\bQ\eta\|}P^c\otimes Q\eta
$$
Therefore, since $\cP_{\eta_Q}(\bP)=\langle\eta_Q|\bP|\eta_Q\rangle$ and $\cP_{\eta_Q}(\bP^c)=\langle\eta_Q|\bP^c|\eta_Q\rangle,$ from the above we obtain that:

\begin{equation}\label{spook1}
\begin{aligned}
\cP_{\eta}(\bP|\bQ)=\cP_{\eta_Q}(\bP)=\langle\eta_Q|P|\eta_Q\rangle = \frac{\langle\eta|\bP\bP^c\bQ|\eta\rangle}{\langle\eta|\bP^c\bQ|\eta\langle}=0,\\
\cP_{\eta}(\bP^c|\bQ)=\cP_{\eta_Q}(\bP^c)=\langle\eta_Q|\bP^c|\eta_Q\rangle=\frac{\langle\eta|\bP^c\bP^c\bQ|\eta\rangle}{\langle\eta|\bP^c\bQ|\eta\rangle}=1.
\end{aligned}
\end{equation}
The equality between the two first terms in each line, stresses the fact that the quantum conditional probability coincides with the probability computed with the von Neumann post-measurement state.

To sum up, if we prepare a system in the special state \eqref{spc1}, our prediction is that if we observe that in the second subsystem, the event $\bQ$ occurs, then in the first subsystem $\bP$ can not occur whereas $\bP^c$ will occur with total certainty. 

Similarly, if we first observe that $\bP$ occurs in the first subsystem, then $\bQ^c$ must occur with probability 1 in the second subsystem. This knowledge acquisition without observation seems to be what Einstein called spooky action at a distance. A simple minded classical analogy goes as follows. Toss a standard die. If the face that comes up is a $1$ you know (without measuring) that the opposite face is a $6.$ The prior knowledge is that the die is a standard die. This is equivalent to the knowledge of the state in which the system was prepared.


\section{The classical case}
Let us consider a very simple case that mimics the quantum example: Two binary random variables taking values in $\{0,1\}.$ The blocks of the probabilistic model are: the sample space $\Omega=\{0,1\}^2.$ The possible events are all subsets of (the sample space) $\Omega,$ and a probability distribution $\cP(i,j)$ from which the probability of any event $A$ can be obtained as $\cP(A)=\sum_{(i,j)\in A}\cP(i,j).$

The analog of the quantum observables are the random variables, which are real valued functions defined on $\Omega.$  We are interested in $X$ and $Y$ defined by $X(i,j)=i$ and $Y(i,j)=j.$ Think of them as coordinate maps if you prefer. From the probability $\cP$ on the events, one obtains, for example,
\begin{equation}\label{pro1}
\cP\big((X,Y)\in A\subset \Omega\big) = \sum_{(i,j)\in A}\cP(i,j) = \cP(A).
\end{equation}
Observe that, for example, since $\{X=i\}=\{(X,Y)\in\{i\}\times\{0,1\}\},$ then $\cP(\{X=i\})=\cP(i,0)+\cP(i,1).$ Table \ref{tab} conveys this notion:
\begin{table}[h!]
\centering\begin{tabular}{|c|cc|c|}\hline
$X\backslash Y$   &  0 & 1 & Row sum \\\hline
0  & $\cP(0,0)$ &  $\cP(0,1)$ & $\cP(X=0)$\\
1  & $\cP(1,0)$ &  $\cP(1,1)$ & $\cP(X=1)$\\\hline
Column sum   & $\cP(Y=0)$ &  $\cP(Y=1)$ & $\cP(\Omega)=1$\\\hline
\end{tabular}
\caption{Joint probability distribution}
\label{tab}
\end{table}
The rightmost column and the last row contain the probabilities of observing the individual values of the variables. A glance at the Table \ref{tab} explains the reduction of the sample space after an observation. Once we know that $\{Y=1\}$ has occurred, the sample space for the next possible predictions changes, and the probabilities have to be recalculated. 

If we think of the sample space as of the microscopic configurations of the system, we may rephrase the former as saying that, even though the number of microscopic configurations in each cell stays the same, since new sample space is smaller (it consists of the configurations compatible with $\{Y=1\}$), we must recompute the new probability of occupation of the cells. The new probabilities are: 
\begin{equation}\label{condpro}
\cP(i|1) = \frac{\cP(i,1)}{\cP(\{Y=1\})}, \;\;\;i=0,1.
\end{equation}
These are the conditional probabilities that must be used to make any further predictions about the system. These are the classical analogue of \eqref{condp1}-\eqref{condp2}.
\section{Final remarks: On entanglement and dependence}
Had we started with an initial state like $\eta=\psi\otimes\phi,$ we would have obtained that 
\begin{gather}
\langle\eta_Q,\bP\eta_Q\rangle = \cP_\eta(\bP)\;\;\;\; \label{corresp1}\\
\langle\eta_Q,\bP^c\eta_Q\rangle = \cP_\eta(\bP^c)\;\;\; \label{corresp2}
\end{gather}
which we interpret as saying that the observation of $\bQ$ does not affect the probabilities of occurrence (or non-occurrence) of the event $\bP.$ Also
\begin{equation}\label{qind}
\cP_\eta(\bP\bQ) = \cP_\eta(\bP)\cP_\eta(\bQ)=\langle\psi,\bP\psi\rangle\langle\phi,\bQ\phi\rangle = \langle\eta,\bP\eta\rangle\langle\eta,\bQ\eta\rangle.
\end{equation}
That is, the probability of the event $\bP\bQ$ is the product of the probabilities of observing the two events. A similar expression is valid replacing either $\bP$ by $\bP^c$ or $\bQ$ by $\bQ^c.$ Observe that the left hand side of \eqref{qind} is independent of the ordering of $\bP$ and $\bQ,$ because $\bP$ and $\bQ$ commute since they act on different factors of the product space. 

Up to this point the analogy with the classical setup is clear, and we may consider the following definition: The classes $\{\bP,\bP^c,\b0,\bbI\}$ and $\{\bQ,\bQ^c,\b0,\bbI\}$ are independent with respect to a state $\phi$ whenever \eqref{qind} holds for any two members in the classes. 

In classical probability, the analogue of the last comments goes as follows.
 The classes $\{\{X=i\}, \emptyset,\Omega\}$ and $\{\{Y=i\}, \emptyset,\Omega\},$ are independent with respect to the probability $P,$ whenever $P(X=i,Y=j)=P(X=i)P(Y=j)$ for any $(i,j)\in\{0,1\}^2.$ In this case, the specification of $Y$ does not change the probability distribution of $X.$

We end up stressing a conceptual issue: Statistical independence implies zero correlation, but the converse may not be true. That is, two random variables might be uncorrelated, but dependent. This is homework for students of basic probability. A simple quantum analogue may go as follows. Consider some observable $\bA\in\cL(H)$ of a quantum system with Hilbert space $H,$ written as $\bA=\sum_ja_j|a_j\rangle.$ The $|a_j\rangle$ are the eigenvectors of $\bA.$ For any state $\psi,$ we put $p_\psi(j)=|\langle a_j|\psi\rangle|^2$ for the probability that $\bA$ has a value $a_j$ in $\psi.$ Let now $f(\bA)=\sum_jf(a_j)|a_j\rangle\langle a_j|,$ and suppose that $f(\bA)$ and $\psi$ are such that
$$ \sum_ja_jp_\psi(j)=0; \;\;\;\mbox{and}\;\;\;\sum_ja_jf(a_j)p_\psi(j)=0.$$
Then, 
$$\langle\psi|\big(f(\bA)-\langle f(A)\rangle_\psi\big)\big(\bA-\langle A\rangle_\psi\big)|\psi\rangle = 0.$$
We use the obvious shorthand $\langle f(A)\rangle_\psi=\langle\psi|f(A)|\psi\rangle$ and $\langle A\rangle_\psi=\langle\psi|A|\psi\rangle.$ The above happens, for example, if $\bA$ has eigenvalues of opposite sign occurring with the same probability. Then, the observables $\bA$ and $f(\bA)$ are uncorrelated, but they are clearly not statistically independent.

Our setup can be rapidly extended to the systems with a finite number of levels, in which the Hilbert spaces are finite dimensional and observables correspond to Hermitian matrices. The technicalities for the general case are elaborate, nevertheless the analogy between prediction with the post-measurement state and prediction with quantum conditional probabilities is maintained.
 
The results presented above can be restated in the Heisenberg representation. Instead of working with vectors in a Hilbert space, one would consider density matrices.In this case there is a larger typology of states between the pure states and the entangled states. See \cite{HHH} for a tour of that zoo. Nevertheless, the correspondence between classical prediction using conditional probabilities and quantum prediction using the post-measurement state is the same.
 
We close mentioning \cite{He} and \cite{Ho} for a connection between classical statistical analysis and quantum signal analysis.

\section{Declaration: Conflicts of interest/Competing Interests}
I certify that there is no actual or potential conflict of interest in relation to this article.

\end{document}